*Title:* **Imaging gate-induced molecular melting on a graphene field-effect transistor**

*Authors:*   Franklin Liou[1,2,3], Hsin-Zon Tsai[1,2], Zachary A. H. Goodwin[4,5,6], Andrew S. Aikawa[1,2], Ethan Ha[1], Michael Hu[1], Yiming Yang[1], Kenji Watanabe[7], Takashi Taniguchi[8], Alex Zettl[1,2,3], Johannes Lischner[4], and Michael F. Crommie[1,2,3]∗

*Affiliations:*

[1]Department of Physics, University of California at Berkeley, Berkeley, CA 94720, United States.

[2]Materials Sciences Division, Lawrence Berkeley National Laboratory, Berkeley, CA 94720, United States.

[3]Kavli Energy NanoSciences Institute at the University of California at Berkeley, Berkeley, CA 94720, USA.

[4]Department of Materials, Imperial College London, Prince Consort Rd, London SW7 2BB, UK.

[5]National Graphene Institute, University of Manchester, Booth St. E. Manchester M13 9PL, United Kingdom

[6]School of Physics and Astronomy, University of Manchester, Oxford Road, Manchester M13 9PL, United Kingdom

[7]Research Center for Functional Materials, National Institute for Materials Science, 1-1 Namiki, Tsukuba 305-0044, Japan

[8]International Center for Materials Nanoarchitectonics, National Institute for Materials Science, 1-1 Namiki, Tsukuba 305-0044, Japan

[†]F.L. and H.-Z. T. contributed equally to this paper

★Corresponding author: crommie@berkeley.edu.

*Abstract:*

Solid-liquid phase transitions are fundamental physical processes, but atomically-resolved microscopy has yet to capture both the solid *and* liquid dynamics for such a transition. We have developed a new technique for controlling the melting and freezing of 2D molecular layers on a graphene field-effect transistor (FET) that allows us to image phase transition dynamics via atomically-resolved scanning tunneling microscopy. Back-gate voltages applied to a $F_4TCNQ$-decorated graphene FET induce reversible transitions between a charge-neutral solid phase and a negatively charged liquid phase. Nonequilibrium molecular melting dynamics are visualized by rapidly heating the graphene surface with electrical current and imaging the resulting evolution toward new equilibrium states. An analytical model has been developed that explains the observed equilibrium mixed-state phases based on spectroscopic measurement of both solid and liquid molecular energy levels. Observed non-equilibrium melting dynamics are consistent with Monte Carlo simulations.

*One-Sentence Summary:*

An electrically tunable solid-liquid molecular phase transition has been imaged on a graphene field-effect transistor with single-molecule resolution.

*Main Text:*

Phase transitions reflect the collective thermodynamic behavior of large numbers of particles, but they fundamentally originate from rapid reconfigurations at the single-particle scale. Numerous techniques have been used to image the dynamics of phase transitions, including high-resolution transmission electron microscopy (TEM),[1,2] low-energy electron microscopy (LEEM),[3] scanning tunneling microscopy (STM),[4] and atomic force microscopy (AFM).[5] These techniques have been applied to a variety of different physical systems, such as electrochemical cells,[6] ion-intercalated systems,[7] surface catalysts,[8] two-dimensional materials,[9] and nanocrystals in solution.[10] However, high-resolution imaging of phase transitions at the single-atom/single-molecule level, including *both* the liquid and solid phases, has so far eluded even the most advanced microscopy due to the non-crystalline nature of liquids and their fast dynamics. Recent progress in this direction has been made using a new technique that is able to image two-dimensional (2D) molecular liquids via STM by confining molecular motion to the surface of a graphene field-effect transistor (FET) and using low temperature (4.5K) to rapidly quench dynamics.[11] Here the time evolution of the liquid state is controlled by passing current through the FET to warm it briefly and thereby control the rate of the liquid kinetics. So far this technique has only been utilized to explore 2D molecular liquids[11] and has not been applied to mixed phase solid-liquid systems.

Here we demonstrate the ability to electrically control a 2D solid-liquid phase transition while imaging all constituent particles of both phases at the atomic scale via STM. This was achieved by depositing 2,3,5,6-Tetrafluoro-7,7,8,8-tetracyanoquinodimethane ($F_4TCNQ$) molecules onto clean graphene FETs having back-gate and source-drain electrodes operable at T = 4.5K in an ultrahigh vacuum STM. We find that lowering the Fermi energy ($E_F$) of the FET via electrostatic gating causes molecular F$_4$TCNQ adsorbates to freeze into a solid, quasi-one-dimensional (1D) chain-phase, while raising E$_F$ causes the molecular solid to melt into a 2D liquid phase. Scanning tunneling spectroscopy (STS) measurements reveal that molecules in the solid phase are charge-neutral while molecules in the liquid phase are negatively charged. By applying short pulses of source-drain current to transiently heat the device, we are able to observe the nonequilibrium dynamics of molecules undergoing both melting and freezing processes. We have developed an analytical theoretical framework that explains the equilibrium energetics of this first-order solid-liquid phase transition as a function of gate voltage (including the mixed phase regime), and we have performed Monte Carlo simulations that capture its nonequilibrium melting dynamics.

The key experimental technique that enables these observations is our ability to tune the diffusivity of adsorbed molecules by applying source-drain current (I$_{SD}$) to the graphene FET to briefly heat the surface (a sketch of the device can be seen in Fig. 1(i)). When the sample is heated by I$_{SD}$ (what we term "diffusive conditions"), simultaneous application of a back-gate voltage (V$_G$) reversibly drives the solid-liquid phase transition of F$_4$TCNQ adsorbates. If I$_{SD}$ is applied for a sufficient amount of time, then the surface molecular configuration reaches a mixed-state equilibrium defined by a specific liquid phase molecular density set by V$_G$. If I$_{SD}$ is set to zero before the surface molecules reach equilibrium then the molecular kinetics halt midway, thus allowing intermediate nonequilibrium states to be imaged as the system evolves toward equilibrium. This technique allows movies to be made of molecular evolution through changing equilibrium landscapes, as well as exploration of fast nonequilibrium dynamics as molecules transition from one equilibrium state to another (see SI).

Fig. 1 shows the reversible melting/freezing of a partial monolayer of $F_4TCNQ$ on a graphene FET as it transitions through different equilibrium states as a function of applied $V_G$. For $V_G = -30$ V (Fig. 1(a)) the molecules all lie in a solid chain phase after flowing a current of $I_{SD} = 1$mA through the device for 180 sec (all images and spectroscopy are acquired only after setting $I_{SD}$ to zero to quench molecular motion). Locating regions on the surface that exhibit the solid phase is challenging because it covers only a small fraction of the device surface (< 10%) under typical experimental conditions, which explains why it was not observed previously.[11] Close-up STM and AFM images of the solid phase reveal two similar quasi-1D chain morphologies that we call "linear" and "zigzag" (Fig. 1(j)).

Subsequent raising of the gate voltage to $V_G = 0$ V under diffusive conditions (*i.e.*, by setting $I_{SD} = 1$ mA for 180 sec) causes the molecular solid to melt. This can be seen in Fig. 1(b) which shows isolated $F_4TCNQ$ molecules dotting the surface near the edge of the solid phase in the same area as Fig. 1(a) (the isolated molecules belong to the liquid phase). Fig. 1(b) shows an equilibrium configuration, meaning that the average concentrations of the liquid and solid phases have stopped changing with time under diffusive conditions (the time to reach equilibrium can vary strongly with gate voltage, but is typically < 60 seconds). The images in Figs. 1(c), (d) show the equilibrium configurations of the same region after incrementally raising the gate voltage first to $V_G = 6$V and then to $V_G = 30$V under diffusive conditions. For every step increase in $V_G$ the solid is seen to melt a little more until it is completely liquefied at $V_G = 30$V. Figures 1(e)-1(h) show the same surface region as $V_G$ is decreased back to -30V under identical diffusive conditions. The liquid-solid phase transition is seen to be completely reversible (movies of the freezing/melting processes are shown in supplemental movies S1, S2).

Justification for calling the molecular phase containing isolated molecules a liquid comes from an analysis of the molecular radial distribution function, g(r), and structure factor, S(**q**). Fig.1(k), for example, shows g(r) extracted from a large-area image containing isolated molecules prepared under equilibrium conditions (Fig. S1(b), see SI for additional details). g(r) shows evenly-spaced peaks with a spacing of *a* = 3.84 nm, as expected for the shell-structure of an isotropic liquid[12]. The structure factor seen in the Fig. 1(k) inset is also indicative of an isotropic liquid and shows no evidence of crystal or gas behavior[12].

Understanding the cause of the observed molecular phase transition requires understanding how charge transfers between molecules and graphene under different gating conditions. STS measurements were used to gain insight into this process by separately measuring the local electronic structure of the solid and liquid phases. Fig. 2(a) shows dI/dV spectra measured on an F$_4$TCNQ chain (solid phase) compared to an isolated F$_4$TCNQ molecule (liquid phase) for $V_G$ = -60V (this is the hole-doped graphene regime as shown by the inset electronic structure diagram in Fig. 2a). The bare graphene spectrum for this surface (taken 10 nm away from any molecules) is shown as an inset for reference. A dip in the bare graphene local density of states (LDOS) near V = 0.34 V marks the location of the graphene Dirac point ($E_D$), thus verifying that graphene is in the hole-doped regime for this gate voltage. The gap-like feature at V = 0 ($E_F$) arises from a well-known phonon-assisted inelastic tunneling (IT) effect[13].

The blue curve in Fig. 2(a) shows the dI/dV spectrum for a single, isolated F$_4$TCNQ molecule (SM) in this hole-doped regime. The leading edge of the first peak marks the LUMO energy as discussed in previous work[11,14] ($E_L^{SM}$ = 0.2 eV and is marked by a dashed blue line), while the second peak ($V_b$ ~ 0.4 V) is a phonon satellite arising from intramolecular vibrations.[14] The F$_4$TCNQ LUMO level is unoccupied for this value of $V_G$. The second curve (red) shows the dI/dV spectrum measured with the STM tip held over the end molecule of an F$_4$TCNQ chain (the chain end (CE) as shown in Fig. 1(j)). The CE spectrum is nearly identical to the single molecule spectrum except that E$_L$ is shifted up by 0.06 eV. The third curve (orange) shows the spectrum for a molecule in the middle of a chain (CM) (as shown in Fig. 1(j)). Here E$_L$ is pushed up even further by an additional 0.05 eV. The overall energy-level structure is schematically represented by the inset sketch which shows the energy level alignment of the SM LUMO, the CE LUMO, and the CM LUMO relative to E$_D$ and E$_F$ (the experimental energy levels of the zigzag and linear chains are identical).

This energy-level structure has important consequences for F$_4$TCNQ/graphene solid-liquid phase transitions. For example, suppose that V$_G$ were first set to V$_G$ = -60V (the case shown in Fig. 2(a)) and then slowly increased under diffusive conditions. This would cause E$_F$ to slide to the right and to eventually intersect with E$_L^{SM}$. The first molecules to fill with charge due to the increasing V$_G$ would thus be isolated F$_4$TCNQ molecules. As shown previously,[11] under these conditions E$_F$ becomes *pinned* close to E$_L^{SM}$ and so never reaches the chain orbitals (E$_L^{CE}$ or E$_L^{CM}$) which therefore remain charge-neutral (i.e., unoccupied) for a wide range of V$_G$ values. Increasing V$_G$ while E$_F$ is pinned in this way causes molecules to melt from the neutral solid and to fill with charge, thereby increasing the density of the charged liquid phase (separation between the isolated molecules is explained by Coulomb repulsion).

A useful thermodynamic variable to characterize this process is the total charge density in the molecule-decorated graphene system, -ΔQ (this counts the excess density of electrons). When the molecular chains begin to melt in response to increased V$_G$, -ΔQ exhibits a discontinuous jump when plotted as a function of E$_F$ as shown in Fig. 2(b). -ΔQ here is obtained from the relationship $-\Delta Q = CV_G$ where C is the capacitance per area between the graphene and the gate electrode. $E_F$ and $E_D$ are measured as a function of $V_G$ from STM spectroscopy (by fitting the dI/dV spectrum as shown in the inset to Fig. 2(a)) and the discontinuity in -ΔQ is observed to occur at $E_F - E_D \approx -0.125$ eV. For $E_F - E_D < -0.125$ eV the molecules are all in the charge-neutral chain phase, so any increase in -ΔQ while $E_F$ is in this regime reflects filling of the graphene Dirac band and follows the well-known parabolic dependence of graphene.[15] When $E_F$ reaches the critical value of $E_F - E_D = -0.125$ eV, however, charge begins to flow into the F4TCNQ LUMO states as the chain phase melts to accommodate the additional charge. The molecules have a high quantum capacitance at this energy, so device charge accumulates rapidly in this regime as $E_F$ is increased and exhibits discontinuous behavior as shown in Fig. 2(b). The $E_F$ dependence of -ΔQ in Fig. 2(b) is reminiscent of the temperature dependence of transferred heat in a standard temperature-driven solid-liquid melting transition (such as ice to water) where latent heat must be provided to increase entropy as the solid converts to a liquid. Here $E_F$ is analogous to temperature and the number of excess electrons (-ΔQ) is analogous to entropy, so one can think of "latent charge" as being necessary to induce 2D molecular melting in our devices.[16]

These insights enable us to develop a theoretical model for quantitatively understanding the microscopic energetics of the F4TCNQ/graphene solid-liquid phase transition. We first note that the F4TCNQ molecules and graphene both exchange electrons with the gate which acts as a reservoir. The thermodynamics of such an open system is described by the grand potential. Under our low-temperature experimental conditions (20~30K with current flow) the entropy contribution TΔS to the grand potential is expected to be small, and so we model the grand potential as follows:

$$\Phi = U - E_F N_e . \tag{1}$$

Here U is the total energy and $N_e$ is the total number of electrons relative to a reference state (i.e., the state where all electrons occupy graphene band states with energy $E < E_L^{SM}$ and the molecules are uncharged). Since the LUMO energy of the chains is higher than that of the isolated molecules, we ignore the possibility of the chains becoming charged and assume that electrons occupy either single-molecule LUMO states or graphene Dirac band states. The graphene contribution to the total energy relative to the reference state is denoted by $U_g(E_F) = \int_{E_L}^{E_F} \epsilon\, g(\epsilon) A\, d\epsilon$, where $g(\epsilon)$ is found from the well-known linear band model[17] to be $g(\epsilon) = \frac{2A(E_D - \epsilon)}{\pi \hbar^2 v_F^2}$. If we assume that our system has a total of N molecules that are all in the neutral chain phase, then the molecular energy can be approximated as $U_s(N) \approx -\alpha N$ where $-\alpha$ corresponds to the energy per bond between adjacent molecules. We denote the number of electrons in this pure solid phase as $N_{e,s}$, in which case the grand potential is

$$\Phi_s = U_s(N) + U_g(E_F) - E_F N_{e,s} \qquad (2)$$

On the other hand, if the $N$ molecules are all in the charged liquid phase then the molecules are each charged by one electron in the LUMO and the molecular contribution to the total energy becomes $U_l(N) = E_L N$ (for simplicity we have dropped the superscript "SM" from $E_L$). We denote the number of electrons in this pure liquid phase as $N_{e,l}$, in which case the grand potential is

$$\Phi_l = U_l(N) + U_g(E_F) - E_F N_{e,l} \qquad (3)$$

The critical Fermi level ($E_F^c$) at which the phase transition occurs is determined by setting $\Phi_s = \Phi_l$. At this Fermi level $N_{e,l} - N_{e,s} = N$ since $N$ electrons are needed to charge the molecules, thereby yielding $E_F^c = E_L + \alpha$. For $E_F < E_F^c$ all of the electrons reside in graphene band states and all of the molecules are condensed into solid chains due to the energy gain of bond formation. For $E_F > E_F^c$, on the other hand, all of the molecules are in the charged liquid state. The transition from the solid phase to the liquid phase does *not* occur when $E_F = E_L$ because melting the chains requires extra energy to break the bond between a chain end molecule and its neighbor (i.e., the latent heat of melting). The process of adding a charged, isolated molecule to the liquid phase only becomes energetically favorable when the Fermi level reaches a value equal to $E_L$ *plus* the energy required to break one bond ($\alpha$). This insight allows us to, in principle, experimentally obtain $\alpha$ by comparing the measured value of $E_F^c$ at which the phase transition occurs (which is marked by Fermi level pinning) to spectroscopic measurements of $E_L$. Experimentally we observe $E_F^c$ to be 120±20 meV below the Dirac point energy and $E_L$ to be 140±5 meV below the Dirac point ($E_L$ was determined previously[11]). The difference between these quantities is on the order of our experimental uncertainty, and so we are not yet able to extract an accurate value of $\alpha$ from our data. We are, however, able to place an upper limit on $\alpha$: $\alpha \lesssim 40$ meV (this is reasonably consistent with a DFT-based estimate of $\alpha$ (see SI)).

While the grand potential is continuous at the phase transition, its first derivative with respect to $E_F$ is not. From Eqs. (2) and (3) we see that $\frac{\partial \Phi_l}{\partial E_F}$ and $\frac{\partial \Phi_s}{\partial E_F}$ differ by $N$ at $E_F = E_F^c$, confirming that this is a first-order phase transition and that the "latent charge" required for complete conversion of N molecules in the solid phase to the liquid phase is the charge of $N$ electrons. This is consistent with the experimental discontinuity in -ΔQ seen in Fig. 2(b) which reflects the charge transferred to melted F4TCNQ while $E_F$ is pinned at the critical value.

The preceding discussion has focused on equilibrium conditions of the pure liquid phase ($E_F > E_F^c$) versus pure solid phase ($E_F < E_F^c$), but we are also able to characterize the nonequilibrium solid-liquid (mixed phase) coexistence regime (i.e., unstable excursions from $E_F = E_F^c$) where the proportion of molecules in the chain and liquid phases can be adjusted from one equilibrium state to another (Fig. 3). Fig. 3(a) shows a plot of the experimental liquid phase molecular density ($N_l/A$, where $A$ is the graphene area) versus $V_G - V_0$ where $V_0 = -10V$ is the gate voltage at which isolated molecules *first* appear in STM images. The yellow dots in Fig. 3(a) show that the experimental equilibrium values for $N_l/A$ exhibit a linear dependence on gate voltage. The magenta dots, on the other hand, show experimental nonequilibrium data obtained by changing $V_G$ and $I_{SD}$ in such a way that diffusive conditions do not last long enough for the system to fully equilibrate. Figs. 3(b)-(g) show a full cycle of the system (measured at a single location on the device) as it evolves from one equilibrium configuration to a different one (yellow dots) and then back again by transitioning through a series of intermediate nonequilibrium states (magenta dots).

We start with Fig. 3(b) which shows a patch of the surface that was initially in an equilibrium state at $V_G - V_0 = 60$ V. At this gate voltage a relatively high liquid phase density ($N_l/A = 4.1 \times 10^{12}$ molecules/cm$^2$) coexists with a much lower concentration of the solid phase. The gate voltage was then changed to $V_G - V_0 = 50$ V under *non-diffusive* conditions (i.e., $I_{SD} = 0$) to set a new equilibrium target, but without allowing the system to evolve toward the new target (since the kinetics are quenched by keeping $I_{SD} = 0$). The resulting nonequilibrium configuration is denoted t = 0 and is visually identical to the equilibrium state at $V_G - V_0 = 60$ V. Fig. 3(c) shows the same region after subjecting it to diffusive conditions (by setting $I_{SD} = 1.1$ mA) for $\Delta t = 50$ ms while holding the gate voltage constant at $V_G - V_0 = 50$ V. The solid phase density is seen to increase, but equilibrium is not yet established. Fig. 3(d) shows the same region after allowing it to evolve for an additional 50ms under diffusive conditions while maintaining $V_G - V_0 = 50$ V. The system is now in equilibrium with $N_l/A$ reduced to $3.5 \times 10^{12}$ molecules/cm$^2$ and the solid density correspondingly increased. Figs. 3(e)-(g) show the same process in reverse as $V_G$ is reset to the original value of $V_G - V_0 = 60$ V. The system is observed to evolve back to its original equilibrium configuration after passing through a nonequilibrium (Fig. 3(f)) regime.

The mixed phase solid/liquid configurations observed in Fig. 3 can be understood within our theoretical framework in a straightforward way. To do this we consider the total energy of a mixed phase state containing $N_l$ molecules in the liquid phase and $N - N_l$ molecules in the chain phase given by

$$U(N_l, E_F) = U_l(N_l) + U_s(N - N_l) + U_g(E_F), \qquad (4)$$

where $U_l$, $U_s$, $U_g$, and $N$ are defined the same as for Eqs. (2) and (3). Here $N$ is constant, and $E_F$ is determined by $V_G$ and $N_l$ (see SI). Only $N_l$ remains variable, and its value at equilibrium $N_l^{eq}$ is obtained by minimizing Eq. (4) with respect to $N_l$ (see SI). The resulting expression for $N_l^{eq}$ per unit area is

$$\frac{N_l^{eq}}{A} = CV_G + \frac{|E_D - (E_L + \alpha)|^2}{\pi \hbar^2 v_F^2}, \qquad (5)$$

where $A$ is the area of the graphene capacitor, $E_L$ is the LUMO energy, $E_D$ is the Dirac point energy, and $v_F$ is the Fermi velocity near the Dirac point ($1.1 \times 10^6 \ m/s$). This expression is similar to an expression derived in ref. 11 using a different approach, but the new expression differs in the last term of Eq. (5) which arises due to the energy required to break a bond ($\alpha$), a factor not considered in ref. 11. Eq. (5) is plotted in Fig. 3(a) (white dashed line) and is seen to match the equilibrium data (yellow dots) quite well. The nonequilibrium behavior (magenta dots) can be explained by plotting $U$ from Eq. (4) as a color map depending on both $V_G$ and $N_l$ in Fig. 3(a). The low-energy region of $U(N_l, V_G)$ is seen to correspond to precisely the equilibrium density defined by Eq. (5) (as expected). Excursions from equilibrium, as shown by the magenta dots, thus push the system to higher energy. The energy landscape of Fig. 3(a) is consistent with the experimentally observed tendency of the system to relax back down in energy to the equilibrium configuration.

A more dramatic example of nonequilibrium behavior is shown in Fig. 4 which exhibits the time evolution of a nonequilibrium melting process at the molecular solid-liquid interface. The STM image in Fig. 4(a) shows that the equilibrium configuration (t = 0) at $V_G$ = -20 V exhibits a region of high solid phase density (upper left) and zero liquid phase density everywhere else. The surface was then put into a nonequilibrium state by rapidly changing the gate voltage to $V_G$ = 60V (corresponding to an expected high equilibrium liquid phase concentration). The system was then allowed to evolve under diffusive conditions for $\Delta t$ = 500μs before being quenched and imaged as shown in Fig. 4(b). This nonequilibrium snapshot shows a "wave" of liquid phase molecules emanating from the molecular solid like water from a melting glacier. The width of the liquid layer extends outward from the solid by ~80nm and exhibits an interparticle spacing that is mostly constant. Fig. 4(c) shows the same area after allowing it to evolve under diffusive conditions for another 700μs. The layer of liquid now extends outward from the solid by more than 160nm. A full video of this process can be found in supplemental movie S3.

The theoretical framework discussed up to now is inadequate to model this type of nonequilibrium dynamics. To better understand this melting process we have generalized our overall model to account for: (i) multiple chains, (ii) isolated *uncharged* molecules, and (iii) screened Coulomb interactions between ionized molecules. We have numerically simulated this more complete model using the Monte Carlo method (see SI) to explain the dynamics shown in Figs. 4(a)-(c). An initial configuration was chosen with molecules arranged into chains to mimic the F$_4$TCNQ solids we observe experimentally (Fig. 4(d)). All model parameters were constrained by experiment except for α (for which we only have an upper bound), but our results do not strongly depend on the precise value of α. A fixed number of electrons was added to the system at the start of the calculation to simulate the gating process, and the resulting liquid phase density and $E_F$ value were subsequently determined. Overall, the simulation produced results quite similar to the experiment. For example, isolated molecules were observed to dissociate from chains after only a few Monte Carlo steps and to move towards empty graphene regions (Figs. 4(e), (f)), similar to the flow of molecules observed experimentally in Figs. 4(b), (c).

In conclusion, we have observed a gate-tunable first-order solid-liquid phase transition for $F_4TCNQ$ molecules adsorbed onto the surface of a graphene FET. We are able to control and image the relative abundances of liquid and solid phases for different equilibrium conditions and to directly visualize the nonequilibrium dynamics of this system with single-molecule resolution for *both* the solid and liquid phases. We have developed an analytical model that explains the gate-dependent equilibrium properties of this system with the only unknown parameter being the energy of cohesion of the molecular solid. The techniques described here provide a new method for experimentally extracting this parameter, and our results put an experimental upper bound on it equal to 40meV per molecule. Monte Carlo simulations show reasonable agreement with the highly nonequilibrium kinetics observed in our experiment. The phenomenology observed here should be generalizable to other adsorbate/surface systems that are gate-tunable.

*Acknowledgments:*

We acknowledge the Molecular Foundry for resources used to fabricate and characterize graphene devices for this work. We acknowledge the Imperial College London Research Computing Service (DOI:10.14469/hpc/2232) for the computational resources used in carrying out this work. We thank J. M. Kahk for useful discussions.

*Funding:*

Director, Office of Science, Office of Basic Energy Sciences, Materials Sciences and Engineering Division, of the US Department of Energy under contract no. DE-AC02-05CH11231, Nanomachine program-KC1203 supported fabrication of graphene devices, STM imaging, and STM spectroscopy

National Science Foundation grant DMR-1807233 supported transport characterization of graphene devices

Kavli ENSI Philomathia Graduate Student Fellowship supported FL

Centre for Doctoral Training on Theory and Simulation of Materials at Imperial College London funded by the EPSRC, EP/L015579/1 supported Monte Carlo simulations

Thomas Young Centre under grant number TYC-101 supported DFT calculation of molecular chain orbital energies

EC-FET European Graphene Flagship Core3 Project, EPSRC grants EP/S030719/1 and EP/V007033/1 supported calculation of graphene free energy

Lloyd Register Foundation Nanotechnology Grant supported DFT calculations of inter-molecule bonding energy

Elemental Strategy Initiative conducted by the MEXT, Japan, Grant Number JPMXP0112101001 supported characterization of BN crystals

JSPS KAKENHI Grant Number JP20H00354 supported growth of BN crystals

*Author contributions:*

Conceptualization: FL, H-ZT, AA, MC

Methodology: FL, H-ZT, AA, ZG, JL

Investigation: FL, H-ZT, AA, EH, MH, KW, TT, MC

Visualization: FL, H-ZT, AA, YY

Funding acquisition: JL, MC

Project administration: JL, MC

Supervision: JL, MC

Writing – FL, H-ZT, ZG, JL, MC

Writing – review & editing: FL, H-ZT, ZG, JL, MC

***Competing interests:***

Authors declare that they have no competing interests.

***Data and materials availability:***

All data are available in the main text or the supplementary materials.

***Supplementary Materials***

Materials and Methods

Figs. S1 to S6

Movies S1 to S3

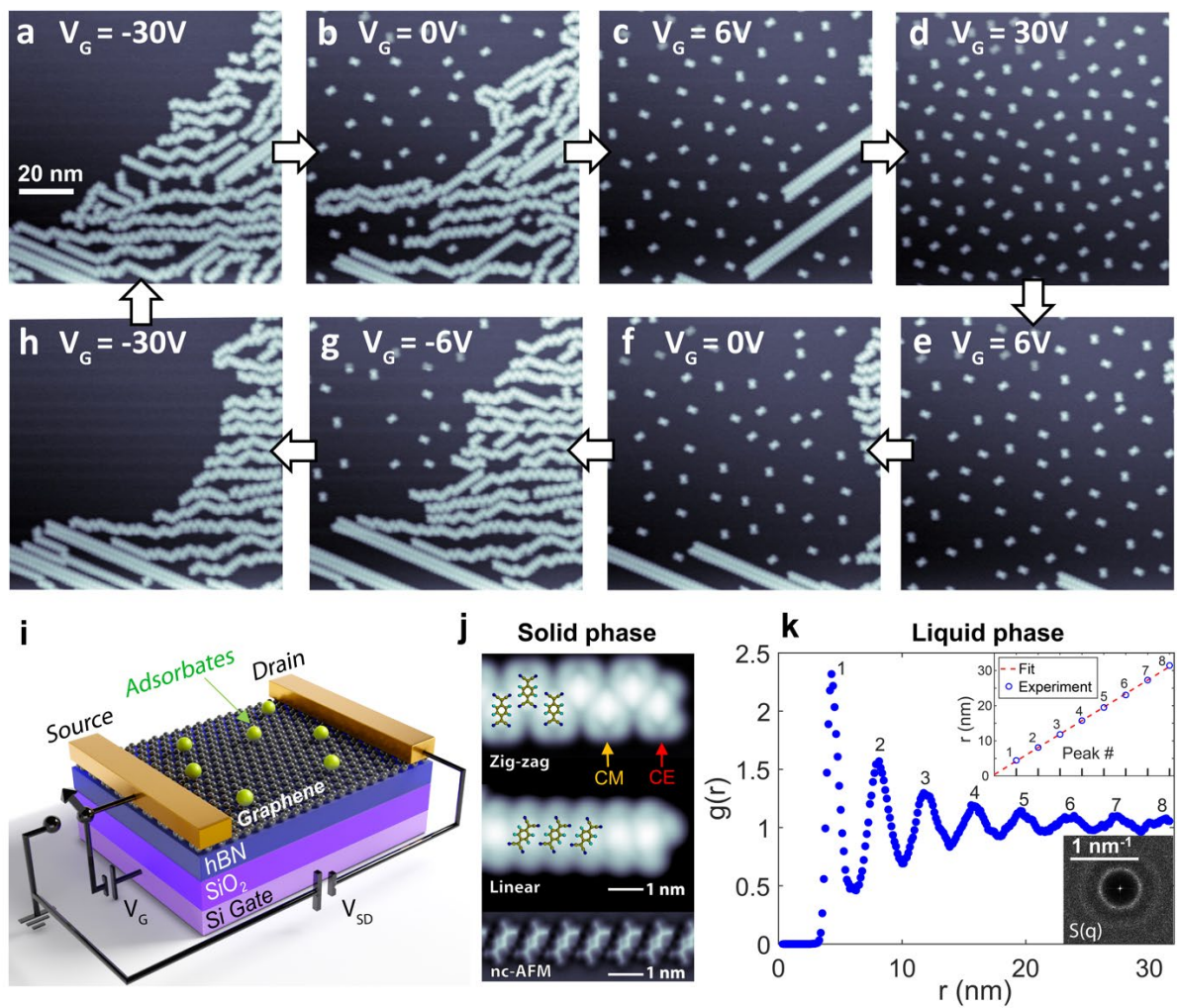

**Fig. 1: Gate-tunable solid-liquid molecular phase transition.** (a)-(d) STM images show the melting of self-assembled chains of $F_4TCNQ$ molecules (solid phase) into isolated molecules (liquid phase) as $V_G$ is increased from -30V, 0V, 6V to 30V. (e)-(h) The reverse phase transition (liquid to solid) is observed at the same spot on the surface with molecules coalescing from liquid phase into self-assembled chains as $V_G$ is decreased to 6V, 0V, -6V and -30V. (i) Schematic of the experimental setup shows $F_4TCNQ$ molecules adsorbed onto the surface of a graphene field-effect transistor (FET) device. (j) Closeup STM images of the molecular chain phase (with structural overlays) show two observed geometries (linear and zig-zag), both having a center-to-center molecular distance of 8.6 Å. A bond-resolved nc-AFM image in the bottom row (obtained with a CO tip[18]) reveals the linear geometry in greater detail. (k) The radial distribution function $g(r)$ of molecular positions in the liquid phase shows a shell-like structure characteristic of liquids having an average shell spacing of 3.84 nm (the corresponding STM image can be seen in Fig. S1(b) of the SI). The corresponding structure factor $S(q)$ shown in the inset indicates that the liquid is isotropic. STM images obtained at T = 4.5K.

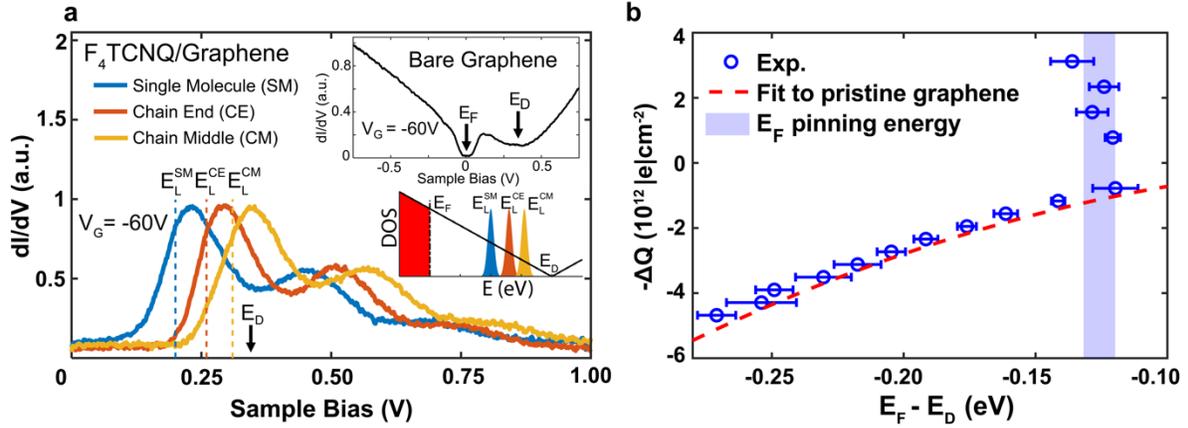

**Fig. 2: Electronic energy level alignment between graphene and F₄TCNQ molecules, and charge accumulation under electrostatic gating.** (a) dI/dV spectra taken at $V_G$ = -60 V for F₄TCNQ molecules at the chain middle (CM), the chain end (CE), and for single, isolated molecules (SM) (images shown in Fig. 1). Inset plot shows dI/dV spectrum measured on bare graphene for $V_G$ = -60V. Inset sketch shows the relative energy alignments of the CM LUMO state, the CE LUMO state, the SM LUMO state, the Fermi energy ($E_F$), and the Dirac point ($E_D$). (b) Total charge density accumulated in the molecule/graphene surface (measured capacitively and plotted in terms of electron density) as a function of $E_F - E_D$ (as determined by STS). The discontinuity at $E_F - E_D$ = -0.125 eV signifies a first-order phase transition. STS spectra obtained at T = 4.5K.

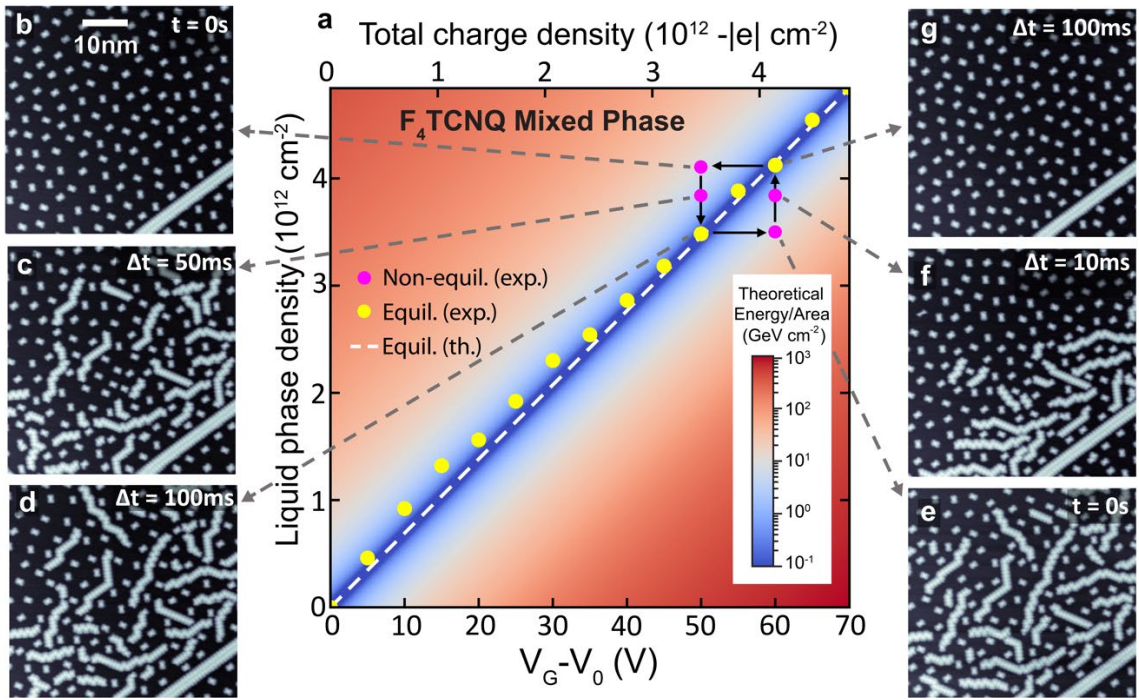

**Fig. 3: F$_4$TCNQ chain freezing and melting under non-equilibrium conditions.**
(a) Experimental values of the equilibrium liquid phase molecule density ($N_l/A$) are plotted as yellow dots and non-equilibrium values as magenta dots. The theoretical total energy of the equilibrium mixed phase of F$_4$TCNQ/graphene is also shown (color scale) as a function of liquid phase surface density and gate voltage ($V_0$ is the gate voltage at which melting first begins). The minimum energy configuration corresponds to the dashed white line (obtained from Eq. (5)). (b) STM image of the non-equilibrium molecular state obtained by switching $V_G - V_0$ to 50 V starting from the equilibrium state at $V_G - V_0 = 60$ V and not allowing the system to evolve under diffusive conditions ($t = 0$). (c) Molecular chains condense into a new non-equilibrium state after allowing the system to evolve for 50ms under diffusive conditions ($I_{SD} = 1.15$ mA, $V_G - V_0 = 50$ V). (d) Molecular chain condensation advances to this equilibrium state after waiting an additional 50ms under diffusive conditions ($I_{SD} = 1.15$ mA, $V_G - V_0 = 50$ V). (e) STM image of the non-equilibrium state obtained by switching $V_G - V_0$ to 60 V and not allowing the system to evolve under diffusive conditions ($t = 0$). (f) Molecular chains have partially melted in this new non-equilibrium state obtained after allowing the system to evolve for 50ms under diffusive conditions ($I_{SD} = 1.11$ mA, $V_G - V_0 = 60$ V). (g) Molecular chains have melted even further in this equilibrium state obtained after waiting an additional 100ms under diffusive conditions ($I_{SD} = 1.11$ mA, $V_G - V_0 = 60$ V), thus returning the molecular density to its initial configuration in (b). STM images obtained at T = 4.5K.

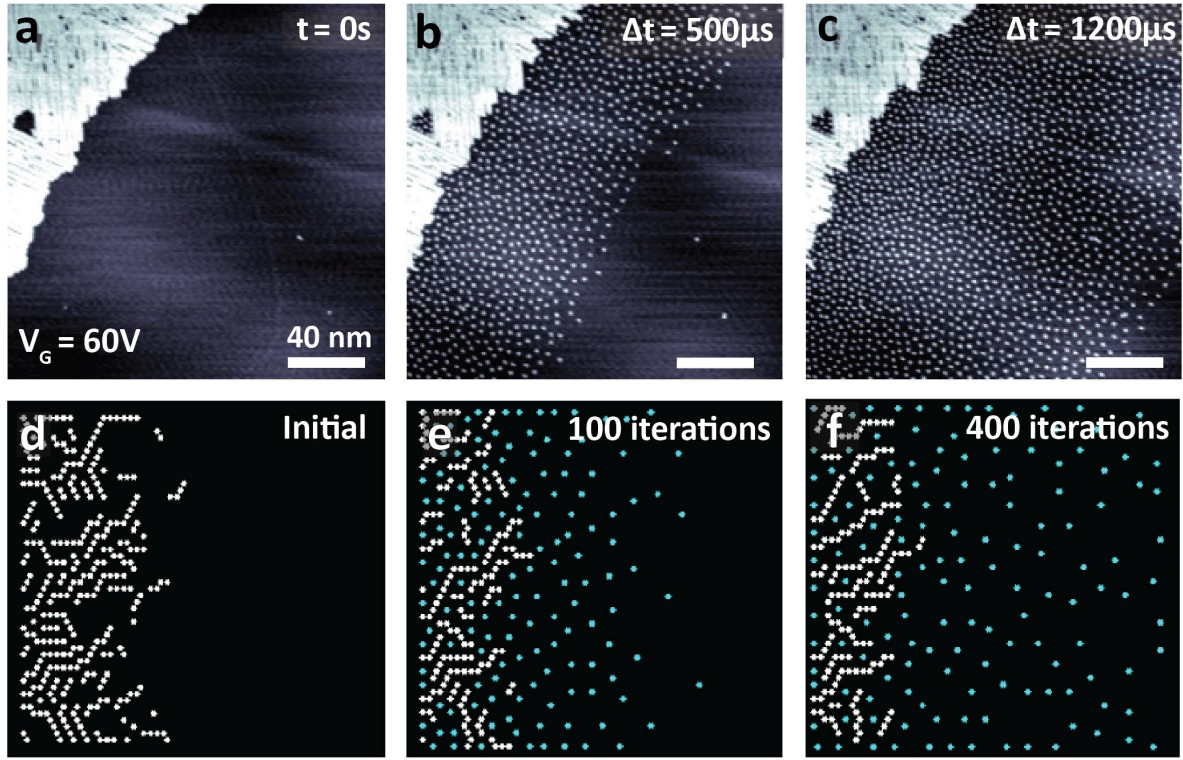

**Fig. 4: Non-equilibrium melting of the F$_4$TCNQ solid.** (a) STM image of equilibrium F$_4$TCNQ solid formed under diffusive conditions on graphene FET at V$_{G-set}$ = -20V. V$_G$ was stepped up to V$_G$ = 60 V before imaging, but the system was not allowed to evolve under diffusive conditions ($t$ = 0). (b) Same region of surface after allowing it to evolve under diffusive conditions for Δt = 500μs (I$_{SD}$ = 1.3 mA, V$_G$ = 60 V). A "wave" of charged liquid phase molecules can be seen emanating from the solid interface. (c) Same region after allowing the system to evolve for an additional Δt = 700μs under diffusive conditions (V$_G$ = 60 V). The flow of the charged molecular liquid has extended even further from the condensed phase interface. (d)-(f) Monte Carlo simulations of F$_4$TCNQ molecules disassociating from chains to model the behavior shown in (a)-(c). Molecules colored in blue are charged and can be seen flowing outward from the charge-neutral condensed phase interface. STM images obtained at T = 4.5K.

# Supplementary Materials
# for

**Imaging gate-induced molecular melting on a graphene field-effect transistor**

Franklin Liou[1,2,3], Hsin-Zon Tsai[1,2], Zachary A. H. Goodwin[4], Andrew S. Aikawa[1,2], Ethan Ha[1], Michael Hu[1], Yiming Yang[1], Kenji Watanabe[5], Takashi Taniguchi[6], Johannes Lischner[4], Alex Zettl[1,2,3], and Michael F. Crommie[1,2,3]*

[1]Department of Physics, University of California at Berkeley, Berkeley, CA 94720, United States.

[2]Materials Sciences Division, Lawrence Berkeley National Laboratory, Berkeley, CA 94720, United States.

[3]Kavli Energy NanoSciences Institute at the University of California at Berkeley, Berkeley, CA 94720, USA.

[4]Department of Materials, Imperial College London, Prince Consort Rd, London SW7 2BB, UK.

[5]Research Center for Functional Materials, National Institute for Materials Science, 1-1 Namiki, Tsukuba 305-0044, Japan

[6]International Center for Materials Nanoarchitectonics, National Institute for Materials Science, 1-1 Namiki, Tsukuba 305-0044, Japan

*Corresponding author: crommie@berkeley.edu.

**This PDF file includes:**

    Materials and Methods
        Graphene transistor fabrication
        STM/STS measurements
        Resolving solid and liquid phase structures with STM
        Device capacitance measurement
        Determining $E_F$ from $V_G$ and $N_l$ (analytical model)
        Determining $N_l^{eq}$ from minimization of $U(N_l, E_F)$
        DFT calculations
        Monte Carlo simulations
    Figs. S1 to S6
    Captions for Movies S1 to S3

**Other Supplementary Materials for this manuscript include the following:**

    Movies S1 to S3

**Materials and Methods**

Graphene transistor fabrication

We fabricated graphene/hBN field-effect transistors (FETs) on highly doped $SiO_2$/Si by mechanical exfoliation. Electrical source and drain contacts were fabricated by depositing 3nm thickness of Cr and 10 nm thickness of Au through a stencil mask. The doped silicon substrate was used as the back-gate. After placing in UHV, the graphene surface was cleaned by high-temperature annealing in vacuum at 400C for 12 hours.

STM/STS measurements

STM/STS measurements were performed under UHV conditions at $T = 4.5$ K using a commercial Omicron LT STM with Pt/Ir tips. STM topography was obtained in constant-current mode. STM tips were calibrated on a Au(111) surface by measuring the Au(111) Shockley surface state before all STS measurements. STS was performed under open feedback conditions by lock-in detection of the tunnel current driven by a wiggle voltage having a magnitude of 6–16 mV rms at 401 Hz added to the tunneling bias. WSxM software was used to process all STM and AFM images.

Resolving solid and liquid phase structures with STM

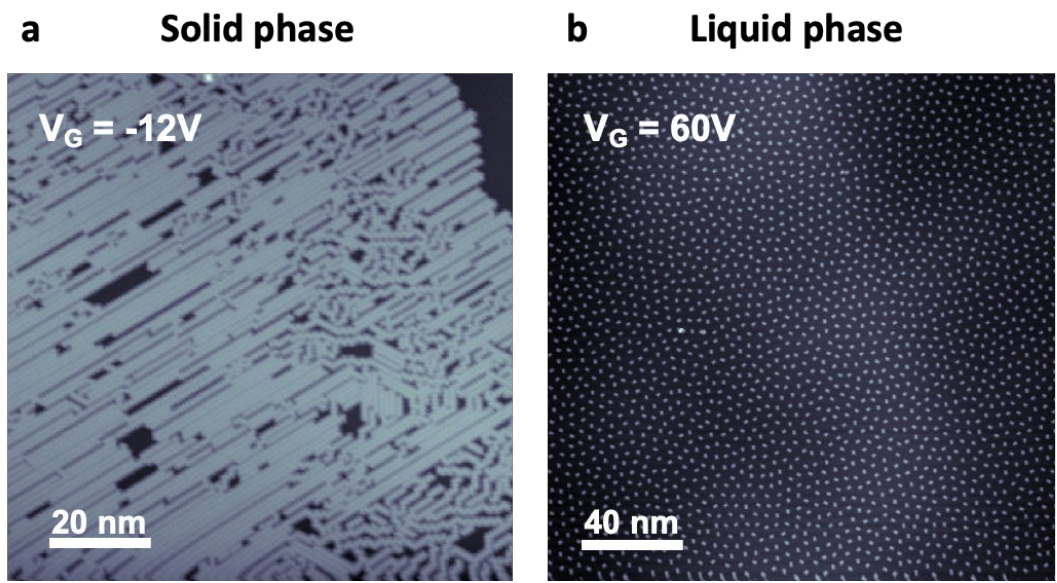

**Fig. S1: Self-assembled F$_4$TCNQ structures on graphene in the solid and liquid phases.** (a) An STM image of molecular chains (the "solid" phase) reveals two coexisting tiling geometries, linear and zig-zag, both with an intermolecular distance of 8.6 Å (I = 1 pA, V = 2V). This image was obtained after allowing the sample to reach equilibrium under diffusive conditions with $I_{SD}$ = 1.9 mA and $V_G$ = -12V. (b) Molecules in the ionic liquid phase show an evenly spaced distribution (I = 1 pA, V = 2V). This image was obtained after allowing the sample to reach equilibrium under diffusive conditions with $I_{SD}$ = 1.15 mA and $V_G$ = 60V. The structure factor $S(q)$ and radial distribution function $g(r)$ plotted in Fig. 1(k) of the main text are both extracted from the molecular positions shown in this image.

Device capacitance measurement

The capacitance of our graphene device was determined by fitting the Dirac point energy as a function of gate voltage obtained from dI/dV spectra taken on pristine graphene. The method we used for fitting the Dirac point energy for each dI/dV curve is described in ref. [11]. We then used the following well-known expression to fit the energy position of the Dirac point as a function of applied gate voltage for a graphene FET:

$$E_D(V_G) = -sgn(V_G)\hbar v_F\sqrt{\pi C|V_G - V'|}, \qquad (S1)$$

where $V'$ is a shift arising from impurity doping (which can vary with location). The extracted value of capacitance per area is $(6.9 \pm 0.1) \times 10^{10}\ |e|\ V^{-1}cm^{-2}$.

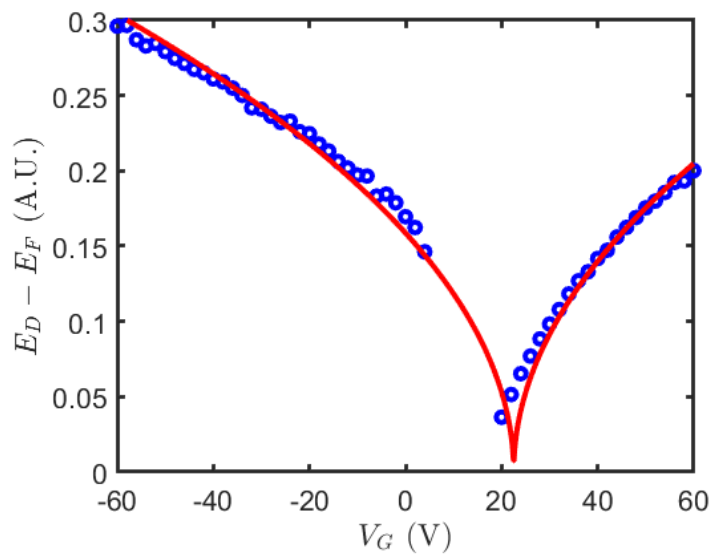

**Fig.S2: Extracting device capacitance:** The device capacitance of $6.9 \times 10^{10}\ |e|\ V^{-1}cm^{-2}$ was extracted by fitting Eq.(S1) (solid curve) to our FET data (circles).

Determining $E_F$ from $V_G$ and $N_l$ (analytical model)

To find $E_F$ (the graphene FET Fermi energy) in terms of $N_l$ (the total number of liquid phase molecules) and $V_G$ (the backgate voltage) we make use of total charge conservation for electrons in the molecular LUMO and graphene band states. Assuming no intrinsic charge doping on the pristine graphene device, the total charge density introduced by electrostatic gating the molecule-decorated device is $-CV_G$ (in units of $|e|\ cm^{-2}$). Electrons introduced by gating can either occupy molecular LUMO states or graphene band states. Since each molecular LUMO can carry one electron of charge, the charge density carried by $N_l$ isolated molecules over an area $A$ is $\frac{-N_l}{A}\ |e|cm^{-2}$. The charge density held in the graphene band states is given by the difference between the Fermi level $E_F$ and the Dirac point energy $E_D$: $\frac{|E_D-E_F|^2}{\pi\hbar^2 v_F^2}$ ($E_F < E_D$ for the regime relevant this work). Combining these charge densities yields:

$$-CV_G = -\frac{N_l}{A} + \frac{|E_D-E_F|^2}{\pi\hbar^2 v_F^2}. \qquad (S2)$$

This expression allows us to express $E_F$ as a function of $N_l$ and $V_G$:

$$E_F = E_D - \sqrt{\pi\hbar^2 v_F^2 \left(\frac{N_l}{A} - CV_G\right)}. \qquad (S3)$$

Determining $N_l^{eq}$ from minimization of $U(N_l, E_F)$:

Starting from the expression for total energy $U$ in Eq. (4) of the main text, we find that for $N$ total molecules on the surface (with the number of liquid phase molecules = $N_l$), $U$ can be expressed as

$$U = E_L N_l - \alpha(N - N_l - 1) + \frac{2A}{3\pi\hbar^2 v_F^2}\left[E_L^3 - E_F^3 + \frac{3}{2}E_D E_F^2 - \frac{3}{2}E_D E_L^2\right]. \qquad (S4)$$

Minimizing $U$ with respect to $N_l$,

$$\left.\frac{\partial U}{\partial N_l}\right|_{N_l=N_l^{eq}} = (E_L + \alpha) + \frac{2A(E_D-E_F)E_F}{\pi\hbar^2 v_F^2}\left.\frac{\partial E_F}{\partial N_l}\right|_{N_l=N_l^{eq}} = 0 \qquad (S5)$$

yields the equilibrium density of molecules $N_l^{eq}/A$ expressed in Eq. (5) of the main text:

$$\frac{N_L^{eq}}{A} = CV_G + \frac{|E_D-(E_L+\alpha)|^2}{\pi \hbar^2 v_F^2} \;.$$

DFT calculations

We performed *ab initio* DFT simulations of molecular chains in vacuum and also for a pair of molecules on graphene using the FHI-aims code (https://doi.org/10.1016/j.cpc.2009.06.022 , https://fhi-aims.org/learn-more) with the PBE functional (https://journals.aps.org/prl/abstract/10.1103/PhysRevLett.77.3865) and a Hirshfeld van der Waals correction (https://journals.aps.org/prl/abstract/10.1103/PhysRevLett.102.073005), as well as a tier 2 basis set for all atoms. These results show that DFT-calculated electronic structure of the molecular chains is consistent with the experimental results shown in the main manuscript, as follows:

**1.** LUMO energies of chains:

In order to understand the chain electronic structure as a function of chain length, the molecules were arranged in zig-zag and linear chains and the atomic positions of molecular chains were allowed to relax until the forces on each atom became smaller than 0.005 eV/Ångstrom, while constraining the *z*-coordinates of all atoms to lie in a plane.

Fig. S3 shows the Kohn-Sham LUMO energies $E_L^{(N)}$ of chains with $N$ molecules (up to $N = 8$) relative to the LUMO energy of an isolated molecule $E_L^{(1)}$, $\Delta E = E_L^{(N)} - E_L^{(1)}$. The LUMO energies of the chains are always higher than the LUMO energy of the isolated molecule and $\Delta E$ approaches approximately 40 meV as $N$ increases. Plotting the wavefunctions of the chain end LUMO states (see Fig. S4) reveals that they are localized on the outermost molecules of the chain. The higher-lying unoccupied states of the chains are formed from the LUMOs of molecules in the middle of the chain. These levels lie approximately 80 meV higher than the single molecule LUMO for long chains. These results are in good agreement with the experimental findings.

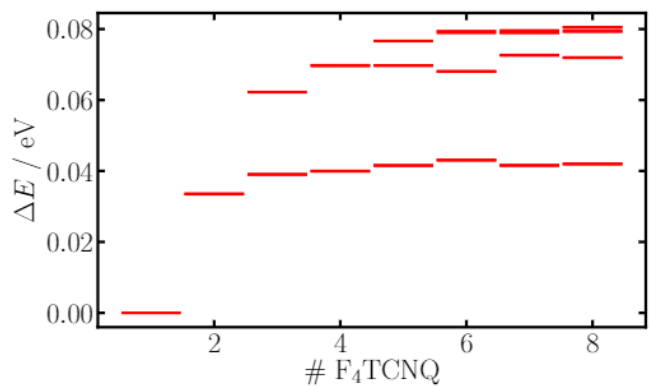

**Fig. S3: F$_4$TCNQ LUMO energies, relative to that of the isolated molecule, obtained from DFT as a function of number of molecules in a molecular chain.** An isolated molecule has the lowest LUMO energy, while molecules within a chain have higher LUMO energies, with the end molecules on the chain having a LUMO which is lower in energy than the rest of the molecules in the chain.

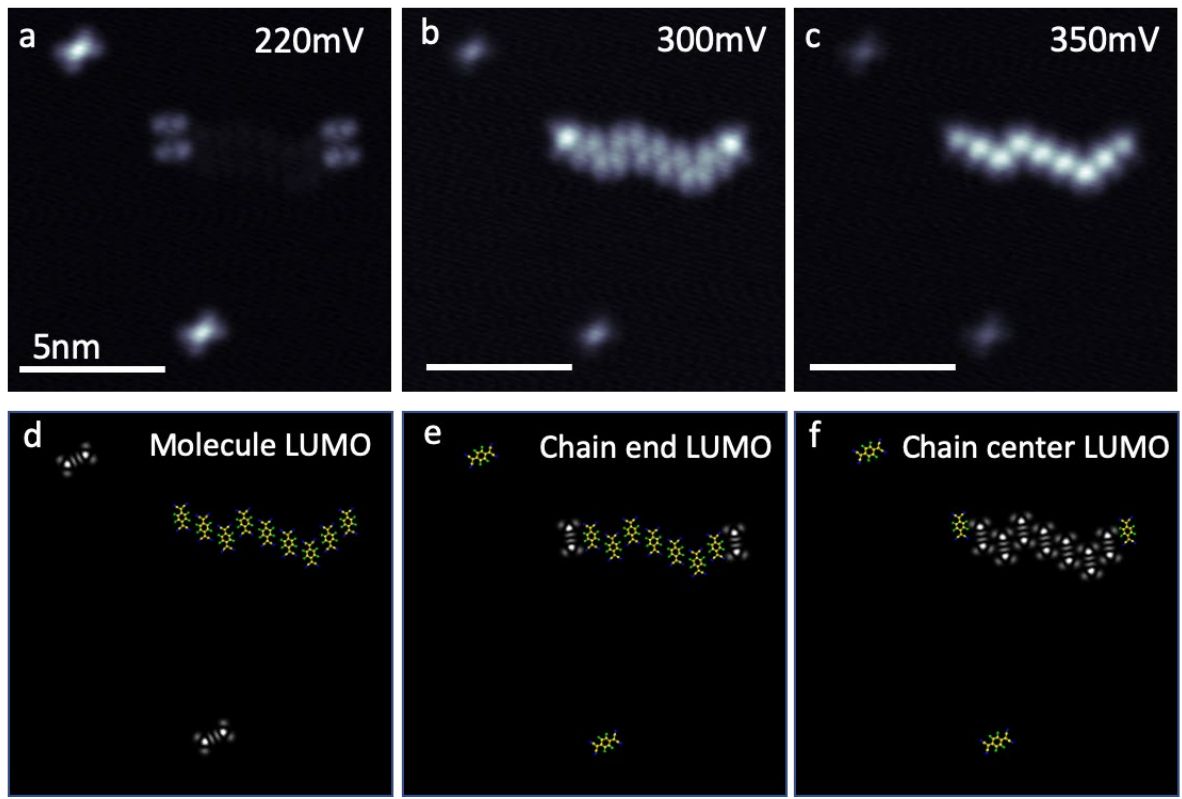

**Fig. S4: dI/dV maps of F$_4$TCNQ near LUMO energies of different molecular configurations.** (a)-(c) Experimental dI/dV maps of a chain and single molecules taken under an external gate voltage $V_G$ = -60V at different energies (for the same region of surface) (a) $V_s$ = 220 mV, (b) $V_s$ = 300 mV, (c) $V_s$ = 350 mV. The dI/dV maps confirm that the single molecule LUMO is the lowest in energy, followed by molecules at the chain end, and then molecules at the center of the chain. (d)-(f) Theoretical electron density contour at energies corresponding to LUMOs of the single molecule, chain end, and chain center.

2. Inter-molecule bonding energy ($\alpha$):

To calculate the bonding energy between F4TCNQ molecules in a chain we prepared a graphene flake consisting of 8x8 graphene unit cells and hydrogen-passivated edges. The structure was relaxed while constraining the z-positions of the atoms to reside in a plane. We then performed constrained DFT calculations for a pair of adsorbed F4TCNQ molecules to ensure that the molecules remain uncharged (reflecting the uncharged character of molecules in chains), see Fig. S5. The binding energy was calculated from

$$-\alpha = E_{pair} - (E_{single,1} + E_{single,2} - E_{gra}), \quad (S6)$$

where $E_{pair}$ is the total energy of both molecules on the graphene flake, $E_{gra}$ is the energy of the graphene flake without any molecules, and $E_{single,1/2}$ denotes the energy when one or the other molecule is removed from the flake. Note $E_{gra}$ is needed to cancel the double counting of the energy of the flake from $E_{single,1} + E_{single,2}$. We find that the bonding energy is $\alpha \sim 45$ meV from these calculations. This is similar to the experimental upper bound on $\alpha$ that was determined from our STM spectroscopy ($\alpha_{exp.} \lesssim 40\ meV$).

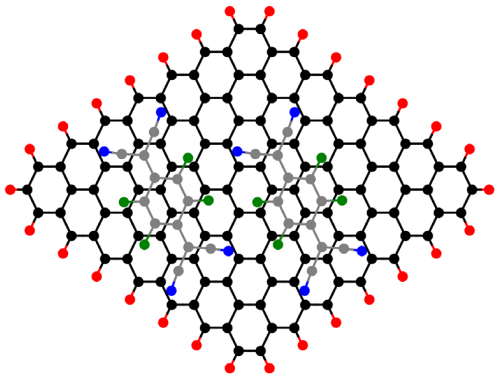

**Fig. S5: Graphene flake with two adsorbed F4TCNQ molecules.**

Monte Carlo simulations

We employed a standard Metropolis Monte Carlo algorithm to simulate the dynamical collective behavior of uncharged molecules in chains together with isolated molecules (both charged and uncharged), and electrons residing in either the LUMO of isolated molecules or in graphene Dirac band states. The thermal energy in the simulations was taken to be $k_B T =1$ meV.

For our simulations the molecules occupy sites on a coarse-grained triangular lattice with an area per site corresponding to 12 graphene unit cells and a lattice spacing of approximately 1 nm, see Fig. S6. A molecule can have one of three possible orientations which point along the vectors connecting a graphene carbon atom to nearest neighbor sites. This lattice is constructed as follows: if a molecule occupies a given site, the neighboring sites (shown in green in the top left panel of Fig. S6) are the ones that the nearest-neighbor molecules would occupy if the molecules were part of the same chain (either zigzag or linear). The red sites in the top left panel of Fig. S6 are nearest-neighbor sites that cannot be occupied because we only consider side-by-side bonding and not end-to-end bonding of molecules (since the latter is not observed experimentally). If a second molecule occupies one of the two neighboring sites on one "side" of the first molecule, then the other site becomes inaccessible (see top right panel of Fig. S6). As a consequence, each molecule can only form up to two bonds with its neighbors. A third molecule can occupy one of two sites on the other "side" of the first molecule which gives rise to either a zigzag arrangement (bottom left panel) or a linear configuration (bottom right panel). The energy per bond is taken to be -10 meV.

Electrons can either occupy graphene states or the LUMO of isolated molecules (we ignore the possibility of charged chains, but allow for uncharged isolated molecules). If a molecule becomes charged, it induces an electrostatic Hartree potential which is experienced by the other charged molecules. The corresponding contribution to the total energy is given by

$$E_H = \tfrac{1}{2} \sum_{i \neq j} W_{ij} n_i n_j \;, \qquad (S7)$$

where $n_{i/j}$ are the charges of the molecules (-|e| or 0), $i$ and $j$ label the isolated molecules located at positions $\tau_{i/j}$, and $W_{ij}$ is the screened Coulomb interaction between these charges. In our simulations, we use the Thomas-Fermi theory result for doped graphene, i.e. $W_{ij} = e^2/(4\pi\epsilon_0 \epsilon \kappa^2 |\tau_j - \tau_i|^3)$ with $\kappa$ denoting the inverse screening length and $\epsilon$ the background dielectric constant. We use $1/(\epsilon\kappa^2) = 25$. In this context the molecules are treated as point charges located on the sites of the effective triangular lattice.

At the beginning of the simulation 300 molecules are distributed randomly on a 50x50 supercell of the effective triangular lattice. The molecular orientations are also initially random. Next, we generate a trial move which is either accepted or rejected by the Monte Carlo algorithm and one iteration of the algorithm is completed when each of the 300 molecules has carried out a trial move. In each move, the molecules can hop from their current site to nearest neighbor sites and also change orientation. We consider three cases: (i) if an isolated molecules hops onto the nearest neighbor site of another molecule, its orientation aligns with that of its neighbor and a bond is formed; (ii) if an isolated molecule hops onto a site without nearest neighbors, its orientation changes randomly; (iii) if a molecule dissociates from a chain, its orientation does not change.

To generate a more realistic initial configuration corresponding to the experimental setup, we performed 160 iterations in the presence of a linear potential in the x-direction with a gradient of -10 meV/nm to mimic the electromigration force of the flowing current used experimentally to create initial molecular configurations. Then 150 electrons were added into graphene states (assuming that all graphene states up to the molecular LUMO level are always filled) and turned off the electromigration potential. In the presence of the electrons, the procedure for a single move in the Monte Carlo simulation was extended according to the following rules: (i) if a molecule is uncharged and not bound to another molecule after it has hopped to a nearest-neighbor site, an electron can be transferred from the highest occupied graphene state to the LUMO; (ii) if an isolated molecule is initially charged and does not form a bond to another molecule in the move, its charge can be transferred to the graphene; (iii) if a molecule is converted from an isolated, uncharged state to a bonded configuration, it cannot become charged in that move; (iv) if a molecule is initially charged and forms a bond to another molecule, its charge must be transferred to the graphene; (v) if a molecule dissociates from a chain, it can become charged.

After each extended move involving a molecular displacement, re-orientation, electron transfer, or bond formation, the change in the total energy of the system (now consisting of a term describing the electrons in the graphene, a term describing the bonding between molecules, a term describing the charging of molecules, and the Hartree interaction between charged molecules) is calculated and the Monte Carlo algorithm decides whether to reject or accept the move. We ran the simulation for 400 iterations.

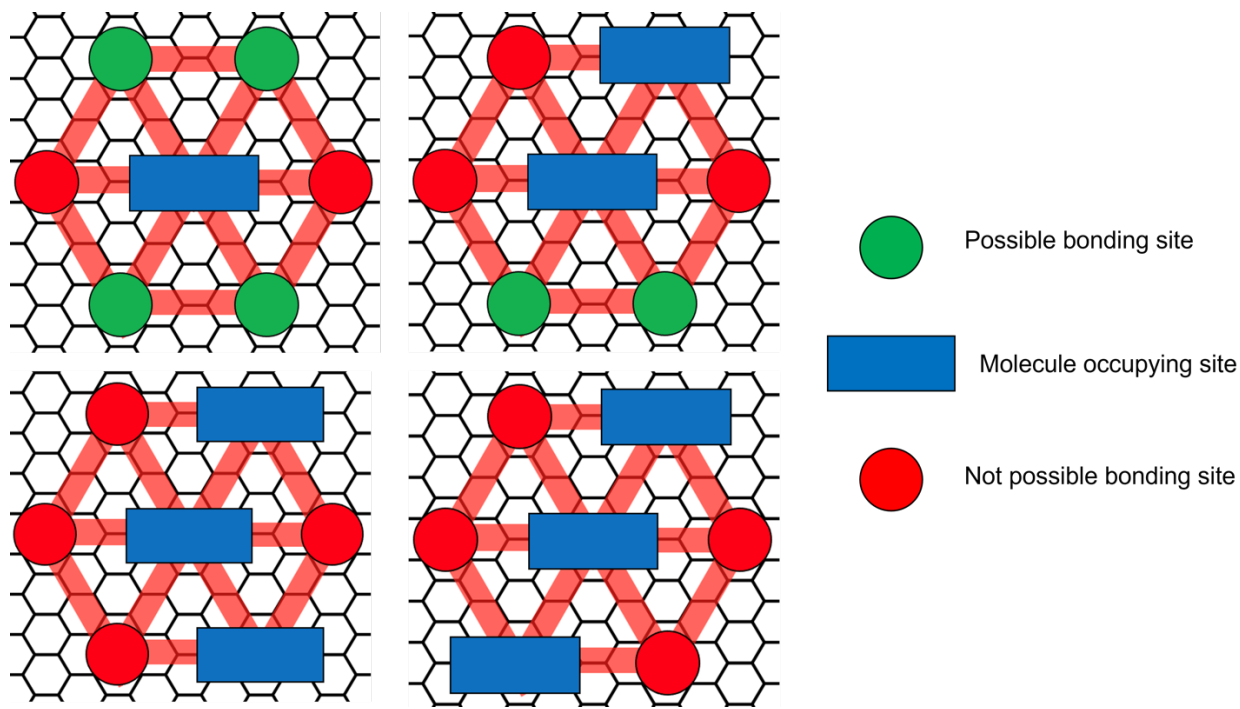

**Fig. S6: Allowed adsorption sites defined in the Monte Carlo simulation:** The allowed adsorption sites form a triangular lattice. Adsorbed molecules can either form a zigzag chain (bottom left) or a linear chain (botton right).

**Movie access:** Three movies showing solid-liquid phase transitions for $F_4TCNQ$ molecules on graphene FET devices (Movie S1, Movie S2, and Movie S3) can be found at this website: https://crommie.berkeley.edu/f4tcnq_movies

**Movie Descriptions:**

**Movie S1: Molecules condensing into molecular chains:** This movie shows sequential STM images of the surface configuration of $F_4TCNQ$ molecules on graphene transforming from a liquid phase to a solid phase (i.e., a freezing transition). At the start of the movie, an equilibrium

molecular configuration is shown that was prepared by simultaneously applying $I_{SD}$ =1.3mA and $V_G$ = 60V to the graphene device for 180s. Each subsequent frame shows the sample after lowering $V_G$ to $V_G$ = -5 V and sending a current pulse through the device for $\Delta t$ = 100 ms (i.e., each frame shows a forward step in time of $\Delta t$ = 100 ms while keeping $V_G$ = - 5 V constant). Molecular chains are observed to continually condense throughout the movie.

**Movie S2: Molecular chains dissociating into single molecules:** This movie shows sequential STM images of the F$_4$TCNQ molecular configuration on graphene transforming from a solid phase back to a liquid phase (i.e., a melting transition). This movie begins where Movie S1 ends, with the surface in a solid molecular configuration (i.e., "frozen") that is obtained after holding the gate voltage at $V_G$ = -5 V under diffusive conditions for $\Delta t$ = 500 ms. In this movie $V_G$ is set to $V_G$ = 60 V and each frame shows the surface evolution after an amount of time $\Delta t$ = 500 μs while keeping the source-drain current at $I_{SD}$ = 1.1 mA. Molecules are observed to dissociate from the chains (i.e., to melt) throughout the movie. By the end of this movie the surface molecular configuration has returned to the equilibrium configuration at $V_G$ = 60 V seen at the beginning of Movie S1.

**Movie S3: Wave of single molecules emerging from molecular chains:** This movie shows sequential STM images of a highly nonequilibrium "wave" of liquid phase F$_4$TCNQ molecules on graphene emerging from a molecular solid. At the start of the movie an equilibrium molecular configuration is shown that was prepared by simultaneously applying $I_{SD}$ =1mA and $V_G$ = -20V to the graphene device for 180s. During the movie $V_G$ is set to 60V and each frame shows the time evolution of the surface after subjecting it to a source drain current of $I_{SD}$ = 1.3 mA for $\Delta t$ = 100 μs. This is a nonequilibrium melting process.